DA: Blockchain and AI in Financial Data Science

IS DOGECOIN A VIABLE INVESTMENT?
INSIGHTS FROM NETWORK AND BUBBLE EFFECTS

Ruoxin Xiao
Xinyu Ying
Hengxu Li
Kexin Liu
(Equally contributed)

June 2022

Is Dogecoin a Viable Investment? Insights from Network and Bubble Effects
Ruoxin Xiao, Xinyu Ying, Hengxu Li and Kexin Liu
June 2022

## **ABSTRACT**

We find that three factors—Dogecoin network externalities, momentum, and tweet sentiment—capture the time-series expected Dogecoin returns. Dogecoin returns are exposed to Dogecoin network factors. We construct the network factors to capture the user adoption of Dogecoin. Moreover, there is a strong time-series momentum effect, and proxies for investor attention strongly forecast future Dogecoin returns. Lastly, we examine potential underlying mechanisms of the Dogecoin price bubble.

# 1      Introduction

Dogecoin is a peer-to-peer, open-source cryptocurrency created in 2013. Unlike other cryptocurrencies, which have marketed themselves on differentiating technical features, Dogecoin's allure likely stems from its cultural roots as a meme (Albi Nani 2022). The community of Dogecoin users have accomplished some rather odd feats, from sponsoring a National Association for Stock Car Auto Racing (NASCAR) driver (Frum, 2014), to sending the Jamaican bobsledding team to the 2014 Winter Olympics in Sochi, Russia (Abbruzzese, 2014) as an homage to the 1993 movie Cool Runnings and launching initiatives called "SaveDogemas" after hacking breach.

The cultural characteristic of Dogecoin is inseparable from its low price, unlimited supply and rapid transactions. Dogecoin is issued with the price of 0.00863 USD and the capitalization of 600 million in 2014, while a member of cryptocurrency asset class with the price of 0.117 USD and the capitalization of 90 billion in 2022. Although Dogecoin's departure from a strictly financial focus to one that was founded on meme culture contributed to its perception as a "joke" currency in popular business media (Hackett, 2020; Jeong, 2017), it has achieved a price increase of more than 50 times and an increase of 10,000 times the market volume in its evolution, among which Elon Musk plays a very important role in the price fluctuations of Dogecoin. In January 2021, Dogecoin went up over 800% in 24 hours as a result of attention from Reddit users, at least in part encouraged by tech baron Elon Musk and the GameStop short squeeze (Usman W. Chohan 2021). In May 2021, Dogecoin hit its all-time high price of 0.627 USD following a comic strip of Elon Musk lifting Dogecoin to the sky posted by himself.

More notably, Dogecoin's value seemed unaffected by other cryptocurrencies, rising even as the rest of the crypto market collapsed (McGuire, 2013). However, the infrastructure and code for Dogecoin was derived from Litecoin, which attempted to improve on Bitcoin's transaction speed and introduce a new hashing algorithm (Padamavathi and Suresh, 2019). Under the Merge Mining, Litecoin miners would be allowed to mine Dogecoin (which is based on the same hashing algorithm with Litecoin) with no need to put in additional computer power, which means they have more chances to gain a reward for their work without much extra cost. As a result, Litecoin miners would be incentivized to secure Dogecoin's network by mining Litecoin and Dogecoin simultaneously. Extraordinarily, the market volume of Dogecoin exceeded Litecoin in April, 2021, which is hard to imagine especially when Dogecoin's system is dependent on Litecoin.

Several recent papers document empirical facts related to cryptocurrency investments. Liu and Tsyvinski (2021) are the first to comprehensively study valuations of the cryptocurrency market in the aggregate time series. They show that cryptocurrency market returns have low exposures to risk factors of the other markets. However, the aggregate market returns can be predicted by cryptocurrency-specific factors such as time-series momentum and investor attention. This paper studies the time-series of Dogecoin returns, considering Dogecoin network, market, and social media messages. We show that Dogecoin returns are driven and can be predicted by these specific factors.

The paper is organized as follows. Section 2 describes the data. Section 3 examines the Dogecoin network and builds the time-series return predictors. Section 4 investigates potential mechanisms behind the Dogecoin price bubble and social media effects. Finally, Section 5 concludes.

## 2  Data

We collect trading data on Dogecoin and Litecoin available from CoinGecko.com, a leading source of cryptocurrency price and volume data suitable for scholarly use (Tomás, 2022). The price of a cryptoasset is calculated based on the pairings available and collected by CoinGecko.com from various exchanges. The price shown is calculated using a global volume-weighted average price formula.

We use daily close prices to construct weekly coin returns. Our sample includes the two coins from the beginning of 2014 to March 2022. Trading volume data became available in the last week of 2013, and thus our sample period starts from the beginning of 2014.

We use three primary measures to proxy the network effect of user adoption: the number of active addresses, the number of transaction count, and the number of payment count. We obtain data on them from Coinmetrics.io. Google search data series are downloaded from Google.

We now document the main statistical properties of the time series of Dogecoin and Litecoin returns. Figures 1, and 2 show the distribution of market returns and market log returns for the two coins at daily, weekly, and monthly frequencies. Figure 3 plots price movements of Dogecoin and Litecoin having some comovements post 2021, but are not perfectly correlated. Because of a large discrepancy in the magnitude of these prices, in Figure 3 we divide Litecoin price by 1000 to improve exposition. Table 1 compares the properties of Dogecoin returns with those of Litecoin returns.

For a particular coin, its excess market return is constructed as the difference between its market return and the risk-free rate measured as the one-month Treasury bill rate from the Federal Reserve Bank of St. Louis' website. Summary statistics are presented in Panel A of Table 1. During the sample period, the Dogecoin returns have a markedly higher mean (4.10% per week) and standard deviation (0.282) compared to the Litecoin returns (2.00% per week with a SD of 0.148). However, the Sharpe ratios of Dogecoin returns (0.23) are comparable to those of Litecoin (0.17) at the monthly frequency. At the daily and weekly frequencies, the negative Sharpe ratios do not convey any useful meaning.

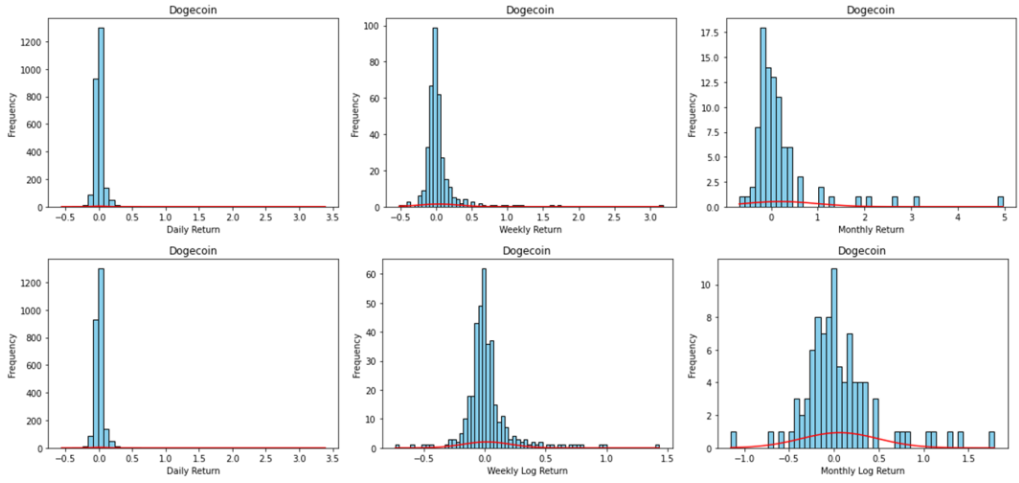

**Figure 1:** Dogecoin Return Distributions

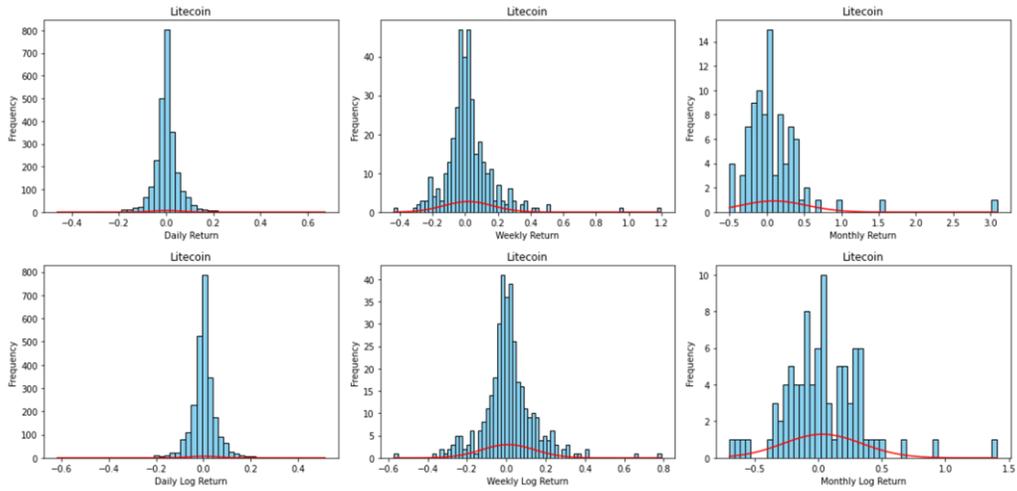

**Figure 2:** Litecoin Return Distributions

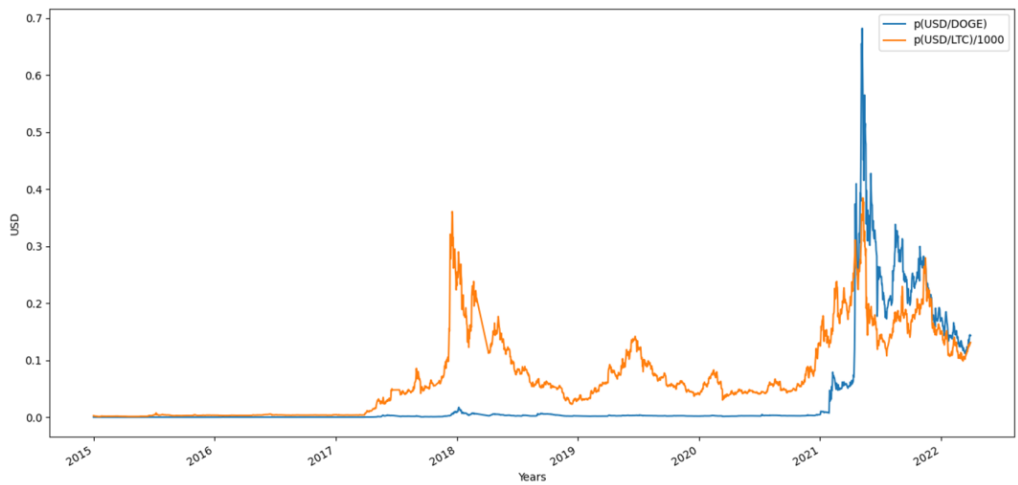

**Figure 3:** Price Movements

Both the Dogecoin and Litecoin returns are positively skewed at all frequencies and have high kurtosis. For Litecoin returns, the skewness increases from 1.13 at the daily frequency to 2.35 at the weekly frequency, and to 3.93 at the monthly frequency, while the skewness of Dogecoin returns has shown the opposite, consistently decreasing from 16.28 at the daily frequency to 3.78 at the monthly frequency.

Both the Dogecoin and Litecoin returns have high probabilities of exceptional negative and positive daily returns. The probability of a -5% Dogecoin daily return is as high as 48.03 percent, quadruple of that of Litecoin.

**Table 1:** Dogecoin and Litecoin Returns Summary Statistics

| Panel A | | | | | | |
|---|---|---|---|---|---|---|
| Daily | Mean | SD | Sharpe | Skewness | Kurtosis | % Return > 0 |
| Dogecoin | 0.59% | 10.04% | - 0.13 | 16.28 | 515.20 | 48.07 |
| Litecoin | 0.33% | 5.99% | - 0.27 | 1.13 | 16.48 | 50.60 |
| | | | | | | |
| Weekly | Mean | SD | Sharpe | Skewness | Kurtosis | % Return > 0 |
| Dogecoin | 4.10% | 28.22% | 0.08 | 5.68 | 49.15 | 45.50 |
| Litecoin | 2.00% | 14.82% | - 0.07 | 2.35 | 15.12 | 52.04 |
| | | | | | | |
| Monthly | Mean | SD | Sharpe | Skewness | Kurtosis | % Return > 0 |
| Dogecoin | 20.00% | 78.71% | 0.23 | 3.78 | 17.48 | 48.91 |
| Litecoin | 9.72% | 44.25% | 0.17 | 3.93 | 23.77 | 54.35 |

| Panel B: Extreme Events of Daily Dogecoin Returns | | | | | | |
|---|---|---|---|---|---|---|
| Disasters | Counts | Percentage | | Miracles | Counts | Percentage |
| < -5 % | 1231 | 48.03 | | > 5 % | 311 | 12.13 |
| < -10 % | 80 | 3.12 | | > 10 % | 150 | 5.85 |
| < -20 % | 14 | 0.55 | | > 20 % | 43 | 1.68 |
| < -30 % | 7 | 0.27 | | > 30 % | 19 | 0.74 |

| Panel C: Extreme Events of Daily Litecoin Returns | | | | | | |
|---|---|---|---|---|---|---|
| Disasters | Counts | Percentage | | Miracles | Counts | Percentage |
| < -5 % | 267 | 10.42 | | > 5 % | 323 | 12.60 |
| < -10 % | 66 | 2.58 | | > 10 % | 111 | 4.33 |
| < -20 % | 9 | 0.35 | | > 20 % | 26 | 1.01 |
| < -30 % | 5 | 0.20 | | > 30 % | 7 | 0.27 |

# 3 Cryptocurrency Factors

The theoretical literature has proposed a number of cryptocurrency-specific factors as drivers of cryptocurrency prices and predictors of cryptocurrency returns. In this section, we develop and investigate the implications of cryptocurrency-specific factors. We first construct Dogecoin network factors. We find that Dogecoin returns are strongly exposed to the Dogecoin network factors. We then test whether Dogecoin returns are predictable by studying whether different cryptocurrency factors predict future Dogecoin returns. We consider momentum and proxies for investor attention. All of these variables are specific to each coin markets. We find that momentum and proxies for investor attention can account for future Dogecoin returns.

### 3.1 Dogecoin Network

The theoretical literature on cryptocurrencies has emphasized the importance of network factors in cryptocurrency valuation (e.g., Cong, Li, and Wang 2019; Sockin and Xiong 2019; Pagnotta and Buraschi 2018; Biais et al. 2018). In particular, the network effect of user adoption can potentially play a central role in the valuation of cryptocurrencies. Because user adoption of cryptocurrencies generates positive network externality, cryptocurrency prices respond to user adoptions. Hence, variations in user adoptions of cryptocurrency networks could contribute to movements in cryptocurrency prices.

We construct network factors of Dogecoin and test whether these factors can account for variations in Dogecoin prices. We use three measures to proxy for network effects: the number of active addresses, the number of transaction count, and the number of payment count. Thus, we measure Dogecoin network growth using the active address growth, transaction count growth, and payment count growth. Panel A of Table 2 reports the correlation across the Dogecoin network factors we consider. The three primary measures correlate with each other positively, with correlations ranging from 0.39 to 0.69.

**Table 2:** Dogecoin and Litecoin Return Loadings to Network Factors

Panel A: Correlation of Network Factors

| DOGE | $\Delta$address | $\Delta$trans | $\Delta$payment | LTC | $\Delta$address | $\Delta$trans | $\Delta$payment |
|---|---|---|---|---|---|---|---|
| $\Delta$address | 1.00 | 0.39 | 0.31 | $\Delta$address | 1.00 | 0.34 | 0.61 |
| $\Delta$trans | | 1.00 | 0.69 | $\Delta$trans | | 1.00 | 0.48 |
| $\Delta$payment | | | 1.00 | $\Delta$payment | | | 1.00 |

Panel B: Network Factor Exposures

| | (1) | (2) | (3) | | (1) | (2) | (3) |
|---|---|---|---|---|---|---|---|
| $\Delta$address | 1.48e-07* (2.15) | | | $\Delta$address | 7.89e-08 (1.61) | | |
| $\Delta$trans | | 2.46e-06*** (6.01) | | $\Delta$trans | | 9.20e-07*** (5.05) | |
| $\Delta$payment | | | 9.61e-07** (2.85) | $\Delta$payment | | | 2.32e-07** (2.63) |
| $R^2$ | 0.0014 | 0.014 | 0.0028 | $R^2$ | 0.00062 | 0.0095 | 0.0023 |

We regress the returns for each coin market on each of the three measures of changes in the coin network. Panel B of Table 2 presents the results using the network factors. The Dogecoin returns positively correlate with all three of the individual Dogecoin network factors. The coefficient on the transaction count, payment count, and active address growth measure is significant at the 0.1%, 1%, and 5% level respectively. Consistent with the theoretical models, these results suggest that the Dogecoin network factors that measure the network effect of user adoptions are important drivers of Dogecoin prices.

## 3.2 Are Dogecoin Returns Predictable?

In this section, we test whether the Dogecoin returns are predictable. Motivated by existing theoretical developments and empirical findings in financial markets, we test whether Dogecoin returns are predictable by momentum and investor attention.

**3.2.1 Cryptocurrency Momentum.** We start by establishing that there is strong evidence of time-series momentum at various time horizons. Panel A in Table 3 documents the Dogecoin time-series momentum results in the regression setting. Specifically, we regress cumulative future Dogecoin returns on current Dogecoin returns from the one-week to eight-week horizons. The current Dogecoin returns positively and statistically significantly predict cumulative future Dogecoin returns at the 0.1% level for all eight horizons. For example, a one-standard-deviation increase in the current Dogecoin return leads to increases in cumulative future Dogecoin returns of 7.06%, 8.75%, 10.72%, and 11.29% increases at the one-week, two-week, three-week, and four-week horizons, respectively. Specifically, the one-week-ahead weekly return is that of buying Dogecoin at 11:59:59 UTD Sunday and selling Dogecoin at 11:59:59 UTD one week later.

**Table 3:** Time-series Momentum

| Panel A: Dogecoin | | | | | | | | |
|---|---|---|---|---|---|---|---|---|
| Weekly | $R_{t,t+1}$ (1) | $R_{t,t+2}$ (2) | $R_{t,t+3}$ (3) | $R_{t,t+4}$ (4) | $R_{t,t+5}$ (5) | $R_{t,t+6}$ (6) | $R_{t,t+7}$ (7) | $R_{t,t+8}$ (8) |
| $R_t$ | 1.39 *** | 1.91 *** | 2.02 *** | 1.89 *** | 2.12 *** | 2.65 *** | 2.73 *** | 2.94 *** |
|  | (17.38) | (19.26) | (14.33) | (9.94) | (7.90) | (8.17) | (6.42) | (6.01) |
| $R^2$ | 0.45 | 0.50 | 0.36 | 0.21 | 0.14 | 0.15 | 0.10 | 0.088 |
| Panel B: Litecoin | | | | | | | | |
| Weekly | $R_{t,t+1}$ (1) | $R_{t,t+2}$ (2) | $R_{t,t+3}$ (3) | $R_{t,t+4}$ (4) | $R_{t,t+5}$ (5) | $R_{t,t+6}$ (6) | $R_{t,t+7}$ (7) | $R_{t,t+8}$ (8) |
| $R_t$ | 0.25 *** | 0.31 *** | 0.38 *** | 0.40 *** | 0.47 *** | 0.56 *** | 0.54 *** | 0.49 ** |
|  | (6.23) | (5.69) | (5.47) | (4.39) | (4.06) | (4.14) | (3.51) | (2.77) |
| $R^2$ | 0.094 | 0.079 | 0.073 | 0.048 | 0.041 | 0.042 | 0.030 | 0.018 |

Additionally, we test whether the time-series momentum effect is linked to network externalities, as suggested in Cong, Li, and Wang (2019). In their dynamic cryptocurrency valuation model, the momentum effect is generated by the positive externality of the network effect that is not incorporated into cryptocurrency prices immediately. That is, their model implies that controlling for cryptocurrency adoption growth would subsume the time-series momentum effect. In Table 4, we show that there is evidence that Dogecoin adoption growth positively predicts future

Dogecoin returns. However, controlling for Dogecoin adoption growth does not subsume the time-series momentum effect documented presented earlier.

**Table 4:** Momentum and Network Effect

| Weekly | $R_{t,t+1}$ (1) | $R_{t,t+2}$ (2) | $R_{t,t+3}$ (3) | $R_{t,t+4}$ (4) | $R_{t,t+5}$ (5) | $R_{t,t+6}$ (6) | $R_{t,t+7}$ (7) | $R_{t,t+8}$ (8) |
|---|---|---|---|---|---|---|---|---|
| $R_t$ | 1.39 *** | 1.91 *** | 2.02 *** | 1.89 *** | 2.12 *** | 2.65 *** | 2.73 *** | 2.95 *** |
|  | (17.36) | (19.22) | (14.30) | (9.92) | (7.89) | (8.16) | (6.41) | (6.02) |
| $\Delta$address | 6.85e-7 | 3.80e-7 | 8.25e-7 | 2.54e-7 | - 1.03e-8 | 2.65e-8 | - 3.25e-7 | - 1.40e-6 |
|  | (1.38) | (0.61) | (0.94) | (0.22) | (- 0.006) | (- 0.021) | (- 0.12) | (- 0.46) |
| $R^2$ | 0.45 | 0.50 | 0.36 | 0.21 | 0.14 | 0.15 | 0.10 | 0.08 |
| $R_t$ | 1.37 *** | 1.91 *** | 1.99 *** | 1.88 *** | 2.11 *** | 2.65 *** | 2.74 *** | 2.95 *** |
|  | (17.20) | (19.07) | (14.11) | (9.85) | (7.82) | (8.13) | (6.40) | (5.99) |
| $\Delta$trans | 1.09e-5 ** | 3.39e-7 | 1.31e-5 * | 2.45e-6 | 3.38e-6 | - 1.04e-6 | - 3.25e-7 | - 2.30e-6 |
|  | (3.30) | (0.81) | (2.24) | (0.31) | (0.30) | (- 0.077) | (- 0.16) | (- 0.11) |
| $R^2$ | 0.47 | 0.50 | 0.37 | 0.21 | 0.14 | 0.15 | 0.10 | 0.09 |
| $R_t$ | 1.38 *** | 1.94 *** | 2.01 *** | 1.91 *** | 2.13 *** | 2.67 ** | 2.81 *** | 3.00 *** |
|  | (17.05) | (19.37) | (14.06) | (9.92) | (7.83) | (8.15) | (6.54) | (6.06) |
| $\Delta$payment | 2.25e-6 | - 6.08e-6 | 2.84e-6 | - 4.26e-6 | - 1.57e-6 | - 5.88e-6 | - 1.8e-5 | - 1.31e-5 |
|  | (0.83) | (- 1.80) | (0.59) | (- 0.66) | (- 0.17) | (- 0.53) | (- 1.25) | (- 0.79) |
| $R^2$ | 0.45 | 0.51 | 0.36 | 0.21 | 0.14 | 0.15 | 0.10 | 0.087 |

**3.2.2 Cryptocurrency Investor Attention.** The theoretical literature of cryptocurrencies has also suggested that investor attention could potentially be linked to future cryptocurrency returns (e.g., Sockin and Xiong 2019). In this section, we investigate the role of investor attention in predicting Dogecoin returns. Specifically, we construct the deviation of Google searches for the word "Dogecoin" in a given week compared with the average of those in the preceding four weeks. We standardize the Google search measure to have a mean of zero and a standard deviation of one. In Panel A of Table 5, we report the results of regressing cumulative future Dogecoin returns from one-week to eight-week horizons on the Google search measure. The Google search measure statistically significantly predicts the one-week to five-week ahead cumulative Dogecoin returns at the 0.1% level. The coefficient estimates of the six-week and eight-week horizons are positive but are less statistically significant. A one-standard-deviation increase in searches leads to increases in weekly returns of about 2% for the one-week ahead cumulative Dogecoin returns and about 2.5% for the two week-ahead cumulative Dogecoin returns.

In Panel B of Table 4, we also report results for Litecoin. Despite being by far the second largest cryptocurrency in terms of capitalization, the current Litecoin returns positively but insignificantly predict week-ahead returns.

**Table 5:** Google Searches

| Panel A: Dogecoin | | | | | | | | |
|---|---|---|---|---|---|---|---|---|
| Weekly | $R_{t,t+1}$ (1) | $R_{t,t+2}$ (2) | $R_{t,t+3}$ (3) | $R_{t,t+4}$ (4) | $R_{t,t+5}$ (5) | $R_{t,t+6}$ (6) | $R_{t,t+7}$ (7) | $R_{t,t+8}$ (8) |
| $Google_t$ | 0.20 *** | 0.25 *** | 0.32 *** | 0.29 *** | 0.31 *** | 0.32 ** | 0.31 * | 0.24 |
|  | (7.07) | (6.57) | (6.92) | (4.96) | (3.83) | (3.28) | (2.46) | (1.69) |
| $R^2$ | 0.12 | 0.10 | 0.11 | 0.061 | 0.036 | 0.026 | 0.014 | 0.0051 |
| Panel B: Litecoin | | | | | | | | |
| Weekly | $R_{t,t+1}$ (1) | $R_{t,t+2}$ (2) | $R_{t,t+3}$ (3) | $R_{t,t+4}$ (4) | $R_{t,t+5}$ (5) | $R_{t,t+6}$ (6) | $R_{t,t+7}$ (7) | $R_{t,t+8}$ (8) |
| $Google_t$ | 0.024 * | 0.033 * | 0.044 * | 0.046 | 0.042 | 0.039 | 0.027 | 0.28 |
|  | (2.04) | (2.05) | (2.14) | (1.75) | (1.25) | (1.01) | (0.62) | (0.40) |
| $R^2$ | 0.0087 | 0.0088 | 0.0098 | 0.0056 | 0.0016 | 2.82e-05 | - | - |

**3.2.3 Interaction between Momentum and Attention.** We have shown that there are strong effects of time-series momentum and investor attention in the cryptocurrency market. The equity market research (e.g., Hong, Lim, and Stein 2000; Hou, Xiong, and Peng 2009) shows that there is a strong relationship between momentum and investor attention. It is possible that these two results capture the same underlying phenomenon. For example, Sockin and Xiong (2019) propose a potential channel to generate momentum. In their model, momentum arises because users have incorrect expectations about future prices—a mechanism similar to De Long et al. (1990). Their model suggests that cryptocurrency momentum and investor attention could potentially arise from the same underlying mechanism. The cryptocurrency momentum and investor attention results could also interact with each other. For example, the Dogecoin time-series momentum effect may be weaker at times of high investor attention, because there is little information leakage at times of high investor attention.

**Table 6:** Google Searches and Past Returns

| Weekly | $Google_t$ (1) | $Google_t$ (2) | $Google_t$ (3) | $Google_t$ (4) | $Google_t$ (5) |
|---|---|---|---|---|---|
| $R_t$ | 1.30 *** | 1.22 *** | 1.23 *** | 1.23 *** | 1.32 *** |
|  | (7.50) | (6.99) | (7.01) | (7.21) | (7.47) |
| $R_{t-1}$ |  | 0.48 ** | 0.50 ** | 0.45 * | 0.41 * |
|  |  | (2.76) | (2.82) | (2.56) | (2.31) |
| $R_{t-2}$ |  |  | -0.11 | -0.16 | -0.13 |
|  |  |  | (-0.61) | (-0.93) | (-0.73) |
| $R_{t-3}$ |  |  |  | 0.36 * | 0.43 * |
|  |  |  |  | (2.08) | (2.43) |
| $R_{t-4}$ |  |  |  |  | -0.37 * |
|  |  |  |  |  | (-2.11) |
| $R^2$ | 0.13 | 0.15 | 0.15 | 0.15 | 0.16 |

First, we show that the current investor attention of Dogecoin is indeed associated with current and past market performance. We regress the current deviation in the Google searches on the contemporaneous and the Dogecoin returns of the previous four weeks. Table 6 documents the results. We find that the deviations in Google searches are positively and significantly associated with contemporaneous and the previous week's Dogecoin returns. The Google search measures do not significantly correlate with past Dogecoin returns beyond one week. Intuitively, these results suggest that investor attention is elevated after superior Dogecoin market performance.

We further test the interaction between the time-series momentum and the investor attention phenomena. The results are reported in Table 7. In Table 7, we regress cumulative future Dogecoin returns on current Dogecoin returns and Google search measures. We find that the coefficients to the current Dogecoin returns are significant for all the horizons, and the coefficients to the Google search measures are significant at the one-week and three-week horizons. The magnitudes of the coefficients to the current Dogecoin returns are similar to the standalone estimates, but the ones to the Google search measures drop by one level. For example, the one-week-ahead coefficients under the univariate regressions are 1.39 and 0.20 for the current Dogecoin returns and the Google search measures, respectively, while they are 1.30 and 0.068 under the bivariate regressions. These results show that the investor attention results actually subsume the time-series momentum.

**Table 7:** Interaction between Momentum and Attention

| Weekly | $R_{t,t+1}$ (1) | $R_{t,t+2}$ (2) | $R_{t,t+3}$ (3) | $R_{t,t+4}$ (4) | $R_{t,t+5}$ (5) | $R_{t,t+6}$ (6) | $R_{t,t+7}$ (7) | $R_{t,t+8}$ (8) |
|---|---|---|---|---|---|---|---|---|
| $R_t$ | 1.30 *** | 1.84 *** | 1.85 *** | 1.75 *** | 1.99 *** | 2.58 *** | 2.69 *** | 3.03 *** |
|  | (15.22) | (17.24) | (12.29) | (8.54) | (6.86) | (7.35) | (5.85) | (5.73) |
| $Google_t$ | 0.068 ** | 0.058 | 0.13 ** | 0.11 | 0.10 | 0.056 | 0.032 | - 0.07 |
|  | (2.83) | (1.92) | (3.14) | (1.91) | (1.26) | (0.56) | (0.24) | (- 0.47) |
| $R^2$ | 0.46 | 0.51 | 0.38 | 0.22 | 0.15 | 0.15 | 0.097 | 0.086 |

## 4  Speculative Interest and Sentiment

As argued by Pástor and Veronesi (2003), it takes time for investors to fully learn and understand emerging technologies, which can lead to price bubbles. We have seven years of Dogecoin market return data spanning from 2014 up to today. There is a great deal of uncertainty and learning about Dogecoin as an asset class during the period. Moreover, the uncertainty has increased since the beginning of 2021. The analysis on the volatility of the Dogecoin market returns also supports this assumption.

In this section, we test whether speculation and investor sentiment may be important drivers of Dogecoin prices. We construct three direct measures of Dogecoin investor sentiment, which are the count of neutral, positive, and negative tweets of Dogecoin on Twitter. We test whether explosive periods of Dogecoin price strongly respond to the contemporaneous and expectations of future speculative price growth. We find sound evidence that the explosive periods positively load on the contemporaneous speculative price growth with significant coefficient estimates in a logistic regression setting.

## 4.1  Bubble Detection

The PSY procedure, known as a practical real-time bubble detection method, was proposed by Phillips, Shi and Yu (2015a, b; PSY hereafter) and was originally designed to identify and date stamp explosive periods in asset prices. Now it has been widely applied to financial, commodity, and real estate markets as an early warning alert system for exuberance.

The method is based on an ADF model specification for the fitted regression equation but uses flexible window widths in its implementation to take time-varying dynamics and structural breaks into consideration. PSY (2015a, b) shows that this algorithm is superior to the forward expanding and rolling window algorithms in bubble identification, especially when the sample period contains multiple bubbles. The dynamics of asset prices during crisis periods may be modeled as a random drift martingale process (Phillips and Shi, 2018)

$$\log P_t = -L_t + \log P_{t-1} + u_t$$

in which $P_t$ is the price of cryptocurrency at time $t$, $L_t$ is a random sequence independent of $u_t$. The sequence $L_t$ produces a random drift in the observed price process following an asymmetric scaled uniform distribution. $L_t$ represent a positive scale quantity measuring shock intensity and hence the process exhibits an overall downward trend.

Let $T_c = \lfloor T \cdot r_c \rfloor$ for sample fraction $r_c > 0$. In the two-regime specification, the logarithmic asset price follows the data generating mechanism (DGP),

$$f(x) = \begin{cases} cT^{-\eta} + \log P_{t-1} + u_t, & \text{if } t \in N \equiv [1, T_c] \\ -L_t + \log P_{t-1} + u_t, & \text{if } t \in C \equiv [T_c, T] \end{cases}$$

in which $N$ denotes the normal market and $C$ denotes collapse periods.

We use backwards sup augmented Dicky Fuller (BSADF) statistics to date stamp the possibly multiple explosive and collapse episodes for Dogecoin by using the backwards sup augmented Dicky Fuller (BSADF) statistics. The BSADF test repeatedly estimates the supremum ADF statistic by considering the backwards expanding time windows ending in one day.

Figure 4 provide the starting and ending dates of explosivity episodes and duration in days bases according to PSY analysis. As shown in Table 8, there were 6 explosive episodes containing 22 days in total in Dogecoin Price. The longest explosivity period was 6 days, whereas the shortest episode of explosivity lasted only 1 day. Following Etienne et al. (2014), who defined the minimum duration of a bubble episode to be three days, Dogecoin prices exhibited 3 bubble periods in April and March 2021. Notably, there is no explosivity periods observed before 2021, during which prices were relatively low.

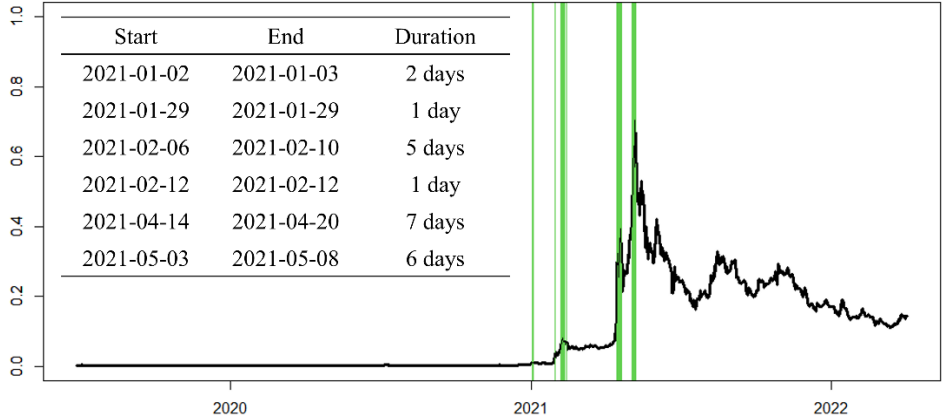

**Figure 4:** Dogecoin Price Bubble and Periods

| Start | End | Duration |
|---|---|---|
| 2021-01-02 | 2021-01-03 | 2 days |
| 2021-01-29 | 2021-01-29 | 1 day |
| 2021-02-06 | 2021-02-10 | 5 days |
| 2021-02-12 | 2021-02-12 | 1 day |
| 2021-04-14 | 2021-04-20 | 7 days |
| 2021-05-03 | 2021-05-08 | 6 days |

## 4.2 Logistic Regression

After examining the explosive periods of Dogecoin prices from 2019 to 2021 through PSY process, we apply logistic regression to see whether the sentiment analysis based on tweets is correlated with the likelihood of explosivity in Dogecoin prices. Let $Y$ be a dummy variable representing the periods of price explosivity, and $X_{i,t}$ represents different kinds of tweets towards Dogecoin, which are neutral, positive, and negative.

$$\log\left(\frac{P(Y=1\mid X)}{1-P(Y=1\mid X)}\right) = \beta_0 + \beta_i X_{i,t} + \varepsilon_t$$

where $Y$ denotes the dummy variable that takes the value of 1 in the presence of explosive periods of price (i.e., if explosivity happens) and 0 otherwise. $\beta_0$ denotes the constant; $X_{i,t}$ is a set of three dummy variables, i.e., one represents neutral tweets, one represents positive tweets and another accounting for negative tweets (2019-2021); and $\varepsilon_t$ is the error term that follows a logistic distribution. The correlation coefficient (i.e. $\beta_i$) between tweets and explosive periods of Dogecoin price is as shown in Table 8.

**Table 8:** Correlation Coefficients for Logistic Regression

| Sentiment | Explosivity |
|---|---|
| Neutral | 1.125 |
| Positive | 2.654 |
| Negative | - 1.921 |

According to the correlation coefficient we obtain from the Logistic Regression Model, neural tweets and positive tweets are positively associated to the period of explosivity while negative tweets have a negative relation. Then we can arrive at a conclusion that Dogecoin-specific tweets do impact the presence of explosivity in Dogecoin price. These results suggest that tweets regarding cryptocurrency can drive explosivity in Dogecoin prices. According to Shiller (2015), price explosivity occurs through psychological contagion where the news regarding the increase in prices spreads contagiously, prompting investors' enthusiasm and bringing a larger group of investors by drawing in with excitement about the past price accelerations. Our results can validate this idea to some extents. This view makes it possible to predict future price dynamics or explosivity based on past information. Moreover, new traders tend to rely on the public than private information like Dogecoin-specific tweets that leads to herding behavior (Banerjee, 1992). The herding behavior can also be caused by some news or reports anticipating certain events, which may be helpful in predicting the price explosivity in the cryptocurrency markets.

## 5    Conclusion

We find that Dogecoin returns strongly respond to Dogecoin network factors, as suggested by the theoretical literature on cryptocurrencies. The returns of cryptocurrency can be predicted by two factors specific to its markets: momentum and investor attention. However, our empirical results do not support the notion that there is only limited interaction between the momentum result and the investor attention result. Moreover, our findings cast doubt on popular explanations that the behavior of Dogecoin returns is driven by its functions as a stake in the future of blockchain technology similar to Litecoin, or as a unit of account similar to other cryptoassets. Finally, we document that the sentiment embodied in social media has a potential to affect Dogecoin prices and lead to a temporary bubble.